\documentclass{article}
\usepackage{graphicx}%
\usepackage{multirow}%
\usepackage{amsmath,amssymb,amsfonts}%
\usepackage{amsthm}%
\usepackage{mathrsfs}%
\usepackage[title]{appendix}%
\usepackage{xcolor}%
\usepackage{textcomp}%
\usepackage{booktabs}%
\usepackage{algorithm}%
\usepackage{algorithmicx}%
\usepackage{algpseudocode}%
\usepackage{listings}%
\usepackage{enumitem}  
\usepackage{float}
\usepackage{hyperref}
\begin{document}

\title{TockyLocus: Quantitative Analysis Methods for Flow Cytometric Fluorescent Timer Data}
\author{Masahiro Ono \\ {\footnotesize \texttt{m.ono@imperial.ac.uk}} \\ {\footnotesize Department of Life Sciences} \\ {\footnotesize Imperial College London} \\ {\footnotesize Imperial College Road, London, SW7 2AZ, United Kingdom}}

\maketitle

\begin{abstract}

Fluorescent Timer proteins, which spontaneously change their emission spectra over time, are valuable tools for analyzing temporal changes in cellular activities at the single-cell level. Traditional analysis of Fluorescent Timer data has mostly relied on conventional flow cytometric methods, which lacks the sophistication needed for detailed quantitative analysis. Recently, we developed the Timer-of-Cell-Kinetics-and-Activity (Tocky) tools, employing transgenic reporter systems using an mCherry mutant Timer protein, Fast-FT, to analyze these changes, implementing data preprocessing methods for Timer fluorescence. Despite this advancement, the computational implementation of effective quantitative analysis methods for Fluorescent Timer data has been lacking. In this study, we introduce rigorous algorithms for a data categorization method, designated as the Tocky locus approach, which uses normalized and trigonometric-transformed Fluorescent Timer data, suitable for quantitative analysis. We found that the five-locus approach, which categorizes Timer fluorescence into five categories, optimally captures the dynamics of Timer profiles, thus enabling effective quantitative statistical analysis and biological interpretation. The current study has developed comprehensive tools for data categorization, visualization, and statistical analysis optimized to Tocky Locus data. These algorithms have been implemented in the \texttt{TockyLocus} package, a new R package designed to facilitate the quantitative analysis of Timer fluorescence data. This toolkit is expected to significantly enhance the utility of experimental systems using Fluorescent Timer proteins, particularly the Tocky tools, by standardizing methods for analyzing dynamic cellular events and responses.

\end{abstract}

\section*{Background}

Fluorescent Timer proteins distinctly exhibit time-dependent changes in their emission spectra within a relatively short period, ranging from several minutes to several hours. This unique characteristic provides critical insights into the temporal dynamics of cellular events. In contrast to GFP and other conventional fluorescent proteins, which typically have long half-lives (e.g., 56 hours for GFP) and primarily reveal accumulated effects, the rapid maturation kinetics of Timer proteins enables the analysis of real-time cellular activities.  Commonly used Fluorescent Timer proteins include DsRed-Timer \cite{Terskikh2000}, mCherry-derived monomeric variants such as Fast-FT, Medium-FT, and Slow-FT \cite{Subach2009}, mRuby variants like mRubyFT \cite{Subach2022}, and 'tandem timers' that combine two fluorescent proteins. These tandem systems often pair Superfolder GFP (sfGFP), a rapidly maturing GFP, with mCherry or TagRFP, making them popular choices for comprehensive Timer studies \cite{Khmelinskii2012, Barry2015}.

Employing Fluorescent Timer proteins as reporter genes enables the unraveling of temporal dynamics in transcriptional activities. By analyzing Nr4a3 transcription downstream of T-cell receptor (TCR) and B-cell receptor (BCR) signaling, temporal changes in activated T and B cells can be elucidated \cite{Bending2018JCB}. This approach also facilitates assessments of promoter activities \cite{Terskikh2000, Troscher2019}. The concept of Timer-of-cell-kinetics-and-activity (Tocky) integrates experimental systems using Fluorescent Timer proteins with transcriptional dynamics analysis, enhancing our understanding of cellular activity over time. Particularly, flow cytometric analysis has proven effective for precisely analyzing Fluorescent Timer profiles, which are essential for deciphering temporal changes in cellular activities and transcription \cite{Bending2018JCB}. 

While the development of Fluorescent Timer transgenic reporter systems is expanding \cite{Bending2018JCB, Bending2018EMBO, Reda2024, Eastman2020, Himuro2024}, a significant challenge remains in the data analysis stage. A key advancement in the Tocky system has been the development of analytical methods to understand Timer fluorescence profiles \cite{Bending2018JCB}. However, the full implementation of these computational algorithms has not yet been achieved. Recently, we introduced the \texttt{TockyPrep} package, which implements essential data preprocessing methods for Fluorescent Timer reporter data \cite{ono2024tockyprep}. Nevertheless, downstream analysis methods, particularly quantitative techniques to analyze nuanced Timer fluorescence dynamics, are crucial. Developing robust computational algorithms for these purposes is essential for establishing the Fluorescent Timer-based methodology to uncover time-dependent changes in individual cells.

In the current study, we introduce a novel R package, \texttt{TockyLocus}, which enables quantitative and statistical analysis of Timer fluorescence profiles using flow cytometric data from Fluorescent Timer reporter systems. The \texttt{TockyLocus} package utilizes the data preprocessing methods implemented in \texttt{TockyPrep} and applies a unique data categorization approach for quantitative analysis.

\section*{Materials and Methods}

\subsection*{Implementation of Tocky Locus Analysis}

To apply Tocky Locus analysis to flow cytometric data, the following analysis pipeline is used:

1. Data preprocessing of Timer fluorescence data with \texttt{TockyPrep} involves:
   \begin{itemize}
     \item \textbf{Thresholding:} Determines Timer fluorescence positivity.
     \item \textbf{Normalization:} Normalizes Timer blue and red fluorescence data.
     \item \textbf{Trigonometric Transformation:} Converts blue and red fluorescence data into Timer Angle and Timer Intensity using polar coordinates.
   \end{itemize}
   
2. Data categorization of Timer Angle data identifies the optimal segmentation into five distinct categories, or Tocky Loci:
   \begin{itemize}
     \item \textbf{0:} New
     \item \textbf{0 -- $<$30:} New-to-Persistent transitioning (NP-t)
     \item \textbf{30 -- $<$60:} Persistent
     \item \textbf{60 -- $<$90:} Persistent-to-Arrested transitioning (PA-t)
     \item \textbf{90:} Arrested
   \end{itemize}

3. Two methods for calculating the distribution of cells within each Tocky Locus are implemented. The default method, the Percentage-Parent, calculates the percentage of cells within each locus relative to the total cell population analyzed. This metric assesses the distribution and predominance of cells across the Tocky Loci within the population of interest. In contrast, the Percentage-Timer calculates the percentage of cells within the population of Timer-positive cells, thereby focusing on proportion analysis. This approach emphasizes the prominent Tocky Loci within Timer+ cells.

3. \textbf{Visualization:} The use of line graphs with individual data points is recommended and implemented to effectively capture nuanced dynamics. This approach allows for the application of statistical methods that are both analytically rigorous and visually interpretable.

4. \textbf{Statistical Application:} Appropriate statistical analyses are applied to Tocky Locus categorized data. Currently, the Mann-Whitney test with p-value adjustment is implemented to ensure robust comparison across categories.

\subsection*{Data Preprocessing Implemented in \texttt{TockyPrep}}

Data preprocessing of Timer fluorescence data was performed using the \texttt{TockyPrep} package. This analysis stage comprises two major processes:

\begin{enumerate}
    \item Timer blue and red fluorescence signals are normalized based on statistics derived from gated negative control cells. Thresholds for red (\( x_{\text{lim.red}} \)) and blue (\( y_{\text{lim.blue}} \)) fluorescence are established either interactively or automatically using quantile-based methods. For each fluorescence channel, the Median Absolute Deviation (MAD) from the log-transformed fluorescence intensities of the gated negative control cells is computed. 
    
    The normalized blue fluorescence (\( B_{\text{norm}} \)) and red fluorescence (\( R_{\text{norm}} \)) for each cell are calculated as follows:

    \begin{equation}
        B_{\text{norm}} = \frac{B_{\text{log}} - \max(B_{\text{log, neg}})}{\text{MAD}(B_{\text{log, neg}})},
    \end{equation}

    \begin{equation}
        R_{\text{norm}} = \frac{R_{\text{log}} - \max(R_{\text{log, neg}})}{\text{MAD}(R_{\text{log, neg}})},
    \end{equation}

    where \( B_{\text{log}} \) and \( R_{\text{log}} \) are the log-transformed blue and red fluorescence intensities of individual cells, and the subscript "\(\text{neg}\)" refers to the negative control cells.

    \item To capture the temporal progression of the Timer fluorescence maturation from blue to red, the normalized fluorescence data are transformed into polar coordinates, yielding two new parameters: Timer Intensity (\( I \)) and Timer Angle (\( \theta \)) \cite{Bending2018JCB}.

    The Timer Intensity \( I \) represents the overall expression level of the Timer protein and is calculated as: $I = \sqrt{B_{\text{norm}}^2 + R_{\text{norm}}^2}$.

    The Timer Angle \( \theta \) reflects the maturation state of the Timer protein, corresponding to the time elapsed since expression. It is computed using the arccosine function:

    \begin{equation}
        \theta = \arccos\left( \frac{B_{\text{norm}}}{I} \right) \times \left( \frac{180}{\pi} \right),
    \end{equation}

Timer Angles range from \( 0^\circ \) (indicative of pure blue fluorescence) to \( 90^\circ \) (indicative of pure red fluorescence).

	\item  Timer Angle data are further analyzed by functions implemented in the \texttt{TockyLocus} package.

\end{enumerate}

\subsection*{Datasets}
\subsubsection*{Nr4a3 Tocky Dataset}
The dataset was derived from flow cytometric analysis of in vitro cultured T cells from Nr4a3-Tocky::OT-II double transgenic mice \cite{Bending2018JCB}. The OT-II transgenic T-cell receptor recognizes the Ova\textsubscript{(323-339)} peptide. Upon TCR signaling, Nr4a3-Tocky T cells express the Fluorescent Timer. Briefly, OT-II Nr4a3-Tocky T cells were activated using Ova (1~\(\mu\)M) in the presence of antigen-presenting cells. Time-course analysis was performed. The Timer fluorescence data from this dataset were partially included in the \texttt{TockyPrep} package. Table~\ref{tab:Table1} shows the group identities of samples within the Nr4a3 Tocky dataset.

\begin{table}[htbp]
\centering
\caption{Summary of Groups, Timer Points, and Treatments}
\label{tab:Table1}
\begin{tabular}{@{}cll@{}}
\toprule
Group & Time (h) & Treatment \\ \midrule
0     & 0        & Stimulation from 0h \\
4     & 4        & Stimulation from 0h \\
8     & 8        & Stimulation from 0h \\
12    & 12       & Stimulation from 0h \\
16    & 16       & Stimulation from 0h \\
24    & 24       & Stimulation from 0h \\
32    & 32       & Stimulation from 0h \\
48    & 48       & Stimulation from 0h \\
32.a  & 32       & Stim. till 24h, then suspended \\
48.a  & 48       & Stim. till 24h, then suspended \\ \bottomrule
\end{tabular}
\end{table}

\subsubsection*{Anti-OX40 Dataset}

This dataset originated from flow cytometric analysis of skin-infiltrating T cells from Foxp3-Tocky mice subjected to allergic skin inflammation and treated with immunotherapy \cite{Bending2018EMBO}. Briefly, the Oxazolone-induced Contact Hypersensitivity model was applied to Foxp3-Tocky mice aged five to ten weeks. Mice were grouped into a control group (isotype rat IgG1 [MAC221] administration) and an anti-OX40 antibody group (0.5 mg anti‐OX40 [OX86] antibody administration), with four mice in the control group and three in the anti-OX40 antibody group. The mice were sensitized with Oxazolone on their abdominal skin on day -5, challenged on their ears on day 0, and analyzed on day 5. Immunotherapy was administered on days 0 and 3 via intraperitoneal injections.

\subsubsection*{Computational Simulation for Protein Maturation Kinetics}

The R package \texttt{deSolve} \cite{deSolve} was used to simulate the kinetics of fluorescent timer proteins through a series of biochemical reactions governed by specific rate constants and initial conditions. Computational simulation was performed using a deterministic model based on ordinary differential equations (ODEs). The model is parameterized by the following rate constants and decay factors, which followed \cite{Subach2009}, \cite{Bending2018JCB}, and \cite{Bending2018EMBO} (expressed in units of per hour):
\begin{itemize}
    \item Rate of conversion from transcript $X$ to intermediate compound $C$: 0.3 per hour
    \item Rate of conversion from compound $C$ to $B$: 8.7 per hour
    \item Rate of conversion from $B$ to $I$: 0.78 per hour
    \item Rate of conversion from $I$ to $R$: 0.14 per hour
    \item Degradation rate of $R$: 0.048 per hour
\end{itemize}

Simulations were run over a predefined time course from 0 to 48 hours, with initial conditions for the transcriptional activity $X$ varied systematically. Time points are defined at 1-hour intervals. Initial values for $X$ range from 100 to 1000.

\paragraph{Parameters:} Simulated Timer blue and red fluorescence data were further processed by logarithmic transformation, normalization and trigonometric transformation using \texttt{TockyPrep} functions\cite{ono2024tockyprep}. 

\subsection*{Software}
The \texttt{TockyLocus} package imports functions from the R packages \texttt{ggplot2}\cite{ggplot2} and \texttt{ggridges}\cite{ggridges2024} for graph production, the package \texttt{RColorBrewer}\cite{RColorBrewer} for generating colors, and \texttt{TockyPrep}\cite{ono2024tockyprep} for data preprocessing of Timer fluorescence data. The Shapiro-wilk normality test was performed using the package \texttt{stats}\cite{R2024}.

\newpage

\section*{Implementation}

\subsection*{Overview}

\begin{figure}[H]
  \centering
  \includegraphics[width=\textwidth]{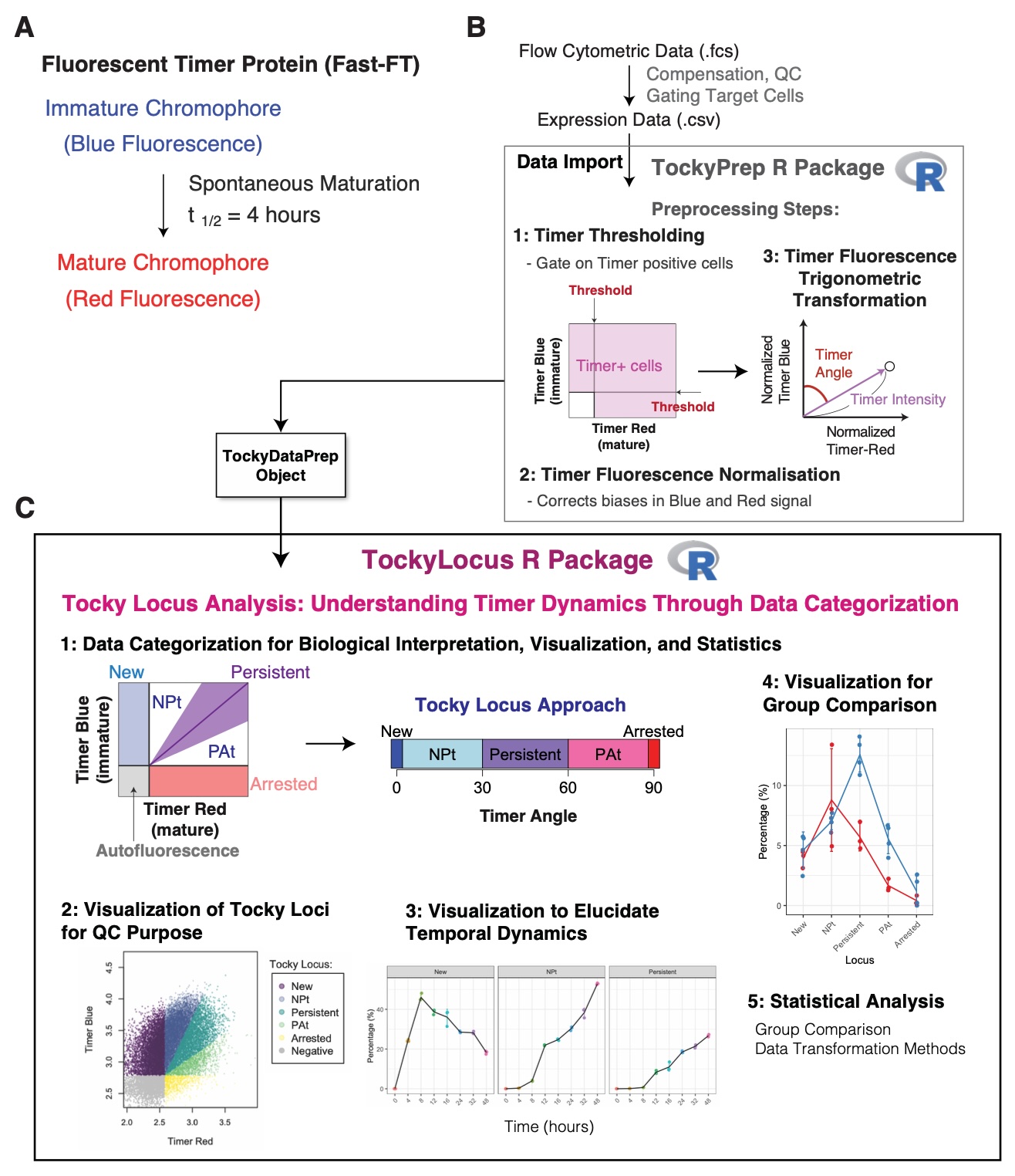}
  \caption{\textbf{Overview of \texttt{TockyLocus} Implementation and its relationship with the \texttt{TockyPrep} package}. (A) The maturation dynamics of Fluorescent Timer protein (Fast-FT). (B) The data import and preprocessing using the \texttt{TockyPrep} package, which applies Timer thresholding, Timer fluorescence normalization, ad Timer fluorescence trigonometric transformation. (C) The preprocessed data is stored as \texttt{TockyDataPrep} S4 object, which is further analyzed using the \texttt{TockyLocus} package, which provides five major functions as depicted.}
  \label{fig:Overview}  
\end{figure}

This section provides the overview for the implementation of Tocky Locus analysis algorithms in the \texttt{TockyLocus} R package (Figure ~\ref{fig:Overview}).

To unravel the dynamics of Fluorescent Timer protein (Figure ~\ref{fig:Overview}A), the \texttt{TockyLocus} package is implemented with unique functions and analyzes preprocessed data using the \texttt{TockyPrep} package. The \texttt{TockyPrep} package, which applies Timer thresholding, Timer fluorescence normalization, and Timer fluorescence trigonometric transformation (Figure~\ref{fig:Overview} B). Note that the preprocessed data is stored as \texttt{TockyDataPrep} S4 object, which is further analyzed using the \texttt{TockyLocus} package.

The \texttt{TockyLocus} package provides five major functions. 

\begin{itemize}
\item(1) Data categirization method using the five Tocky loci
\item(2) Visualization of Tocky Loci in flow cytometric plot for Quality Control (QC) purpose
\item(3) Visualization of temporal dynamics using the locus-wise plots
\item(4) Visualization for group comparisons
\item(5) Statistical analysis methods.
\end{itemize}

\subsection*{Statistical Testing Methods}

Three statistical testing methods are implemented:

\begin{enumerate}
    \item \textbf{Wilcoxon Rank-Sum Test (Mann-Whitney U Test):}
    \begin{itemize}
        \item This non-parametric test is used to compare the distributions of percentages between two groups without assuming normality.
        \item The test is applied to each locus individually.
        \item P-values are calculated, and multiple testing correction is performed using the Benjamini-Hochberg (BH) method or another specified adjustment method.
        \item Significance is determined at an adjusted p-value threshold of 0.05.
    \end{itemize}

    \item \textbf{Arcsine Square Root (ASR) Transformation with T-Test \cite{Nowicka2019, DeBiasi2021}:}
    \begin{itemize}
        \item Percentage data, being proportions bounded between 0 and 100\%, are transformed using the arcsine square root transformation to stabilize variances and approximate normality.
        
        \[
    \text{Transformed\_Percent}  = \arcsin\left(\sqrt{\frac{\text{Percent}}{100}}\right)
        \]

        \item A Shapiro-Wilk normality test is conducted on the transformed data to confirm the appropriateness of parametric testing.
        \item Upon confirmation of normality (p-value $> 0.05$), independent two-sample t-tests are performed for each locus to compare group means.
        \item P-values are adjusted for multiple comparisons using the specified method, and significance is assessed at the 0.05 level.
    \end{itemize}

    \item \textbf{Logit Transformation with T-Test \cite{Nowicka2019, Seiffert2024}:}
    \begin{itemize}
        \item An alternative transformation using the logit function is applied to the percentage data:
        \[
        \text{Transformed\_Percent} = \log\left(\frac{\text{Percent} + c}{100 - \text{Percent} + c}\right),
        \]
        where \( c \) is a small constant (e.g., 0.001) added to handle cases where the percentage is exactly 0 or 100.
        \item Similar to the ASR method, a Shapiro-Wilk normality test is performed on the transformed data.
        \item If normality is confirmed, t-tests are conducted for each locus.
        \item P-values are adjusted, and significance is determined as above.
    \end{itemize}
\end{enumerate}

\paragraph{Multiple Testing Correction}

For all statistical tests, p-values are adjusted to account for multiple comparisons across the five loci. The default method used is the Benjamini-Hochberg procedure, which controls the false discovery rate (FDR). Other adjustment methods available in the \texttt{p.adjust} function in R, such as Holm or Bonferroni corrections, can also be specified.

\newpage

\section*{Results}

\subsection*{Conceptual Definition of Major Tocky Loci by Flow Cytometric Timer Fluorescence Data}
Considering the rapid maturation of Timer blue fluorescence and the stability of Timer red fluorescence, three distinct Timer profiles can be identified within the Timer blue and red fluorescence space \cite{Bending2018JCB}. This subsection aims to clarify the \textit{a priori} definitions within Timer fluorescence data—those determined by logical deduction from the known properties of fluorescent proteins and flow cytometry. These definitions serve as the conceptual framework for identifying distinct loci for Timer dynamics.

\subsubsection*{Definition of Timer Fluorescence Positivity:}
\begin{figure}[H]
  \centering
  \includegraphics[width=0.75 \textwidth]{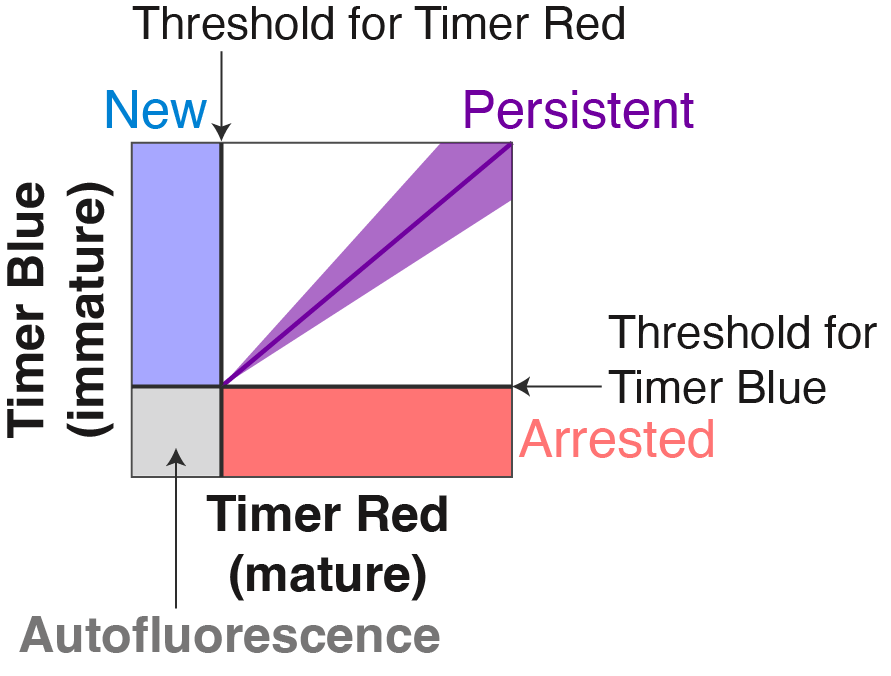}
  \caption{\textbf{Definition of Major Tocky Loci:} (A) Data setup using \texttt{prep\_tocky} and preprocessing steps executed by \texttt{timer\_transform}, which include Timer Thresholding, Timer Fluorescence Normalization, and Trigonometric Transformation.}
  \label{fig:TockyLocus_def}  
\end{figure}

Given the limitations of flow cytometry, signals below a certain threshold level for fluorescence are indistinguishable from autofluorescence, which is inherent to various cellular components including small molecules such as NADPH \cite{Schaefer2019}. The autofluorescence range also includes signal noise \cite{Jameson2022}. To accurately characterize Timer profiles, thresholds for Timer blue and red fluorescence must be established. This step is performed using Timer thresholding in the \texttt{TockyPrep} package (Figure~\ref{fig:TockyLocus_def}).

\subsubsection*{Definition of New, Persistent, and Arrested Loci:}

Following the establishment of fluorescence positivity, cells with or without Timer blue fluorescence are defined as Timer Blue+ and Timer Blue-, respectively. Similarly, cells with or without Timer red fluorescence are categorized as Timer Red+ and Timer Red- \cite{Bending2018JCB}. 

\textbf{New Locus:} Considering that the Timer blue chromophore matures into the red chromophore with a half-life of 4 hours, Timer Blue+ Red- cells indicate that these cells have recently activated Timer transcription and expressed the Timer protein. 

\textbf{Persistent Locus:} Continuous transcriptional activities result in an accumulation of cells towards a Timer Angle of 45 degrees, representing the steady state of transcriptional dynamics \cite{Bending2018JCB}.

\textbf{Arrested Locus:} Timer Blue- Red+ cells are those that have once activated Timer transcription and accumulated some levels of Timer red proteins but have ceased Timer transcription long enough to allow the decay of Timer blue fluorescence. 

\subsection*{Computational Simulation of Timer Dynamics by Constant Transcriptional Activities}

\begin{figure}[H]
  \centering
  \includegraphics[width=\textwidth]{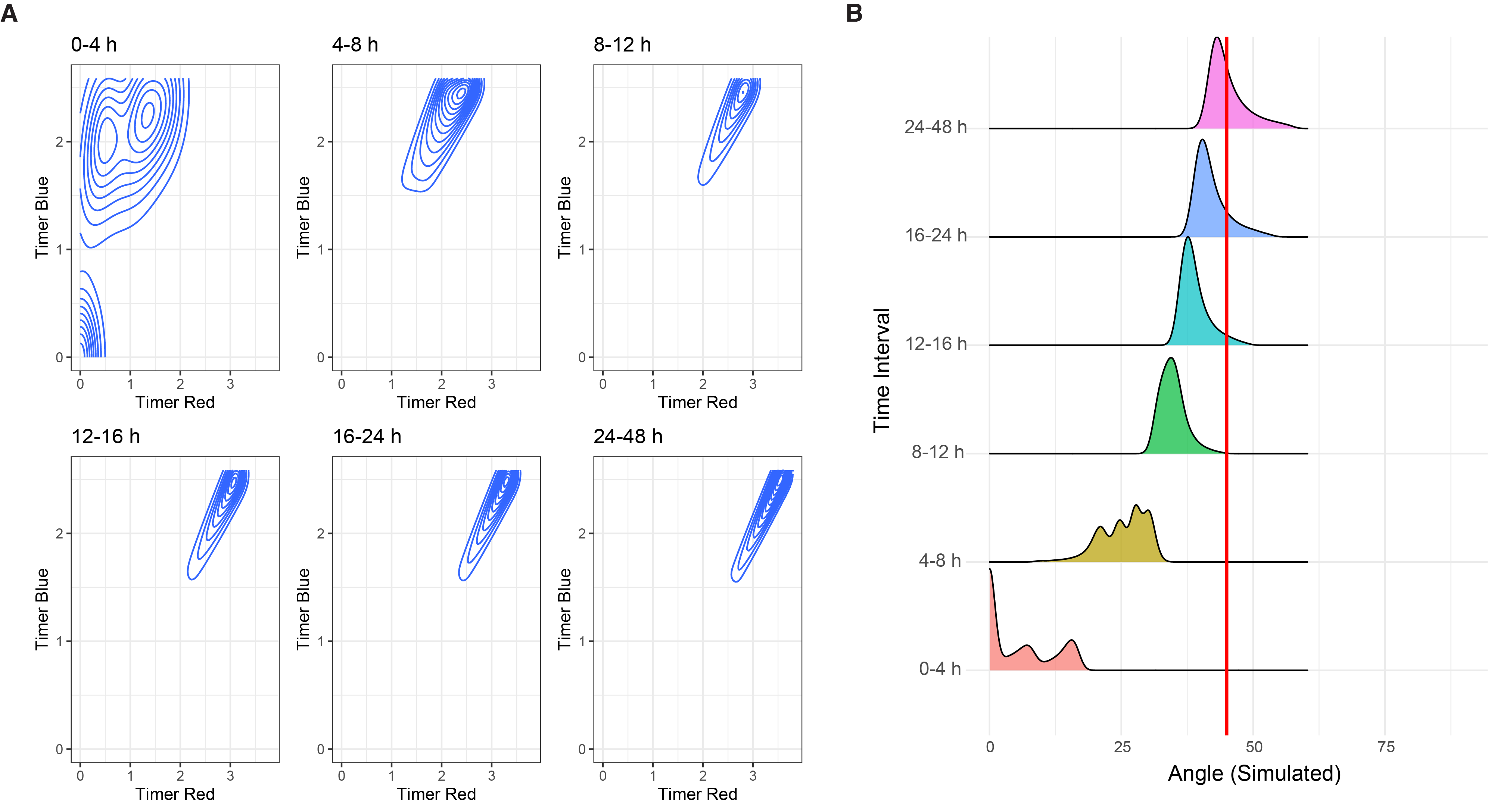}
  \caption{\textbf{Computational Simulation of Timer Dynamics:} (A) Simulated Timer fluorescence dynamics. Contour plots were generated for cells that are within each of the time intervals. See Methods section for the details of the simulation parameters. (B) Progression of Timer Angles post-transcription activation. The data in (A) was used to obtain Timer Angle data using the \texttt{TockyPrep} package. Density plot was generated for the cells that are within each of the time intervals.}
  \label{fig:Fig_sim_constant}  
\end{figure}

To investigate the progression of Timer Angle dynamics under constant transcriptional activities, we performed a series of computational simulations. These simulations aimed to correlate flow cytometric Timer fluorescence with Timer Angle progression (Figure~\ref{fig:Fig_sim_constant}). The analysis captured the previously reported dynamics of Timer fluorescence \cite{Bending2018JCB} (Figure~\ref{fig:Fig_sim_constant}A). Subsequently, the Timer fluorescence data were transformed into Timer Angles using the \texttt{TockyPrep} package \cite{ono2024tockyprep}. The Timer Angle demonstrated a progressive increase following the activation of Timer transcription (Figure~\ref{fig:Fig_sim_constant}B). Most importantly, the cells approached a Timer Angle of \(45^\circ\) as they continuously transcribed the Timer gene.

\subsection*{Experimental Validation of Timer Fluorescence Dynamics}

Using the Nr4a3 Tocky dataset \cite{Bending2018JCB}, we aimed to refine methods for quantitatively analyzing Timer Angle dynamics. Table~\ref{tab:Table1} outlines the group identities and analysis time points. The Timer fluorescence data were converted into Timer Angle and Intensity using the \texttt{TockyPrep} package, which facilitated a detailed assessment of the dynamics involved.

\subsection*{Limitations of Common Approach to Fluorescent Timer Data using Mean Fluorescence Intensities (MFI)}

\begin{figure}[H]
  \centering
  \includegraphics[width=1\textwidth]{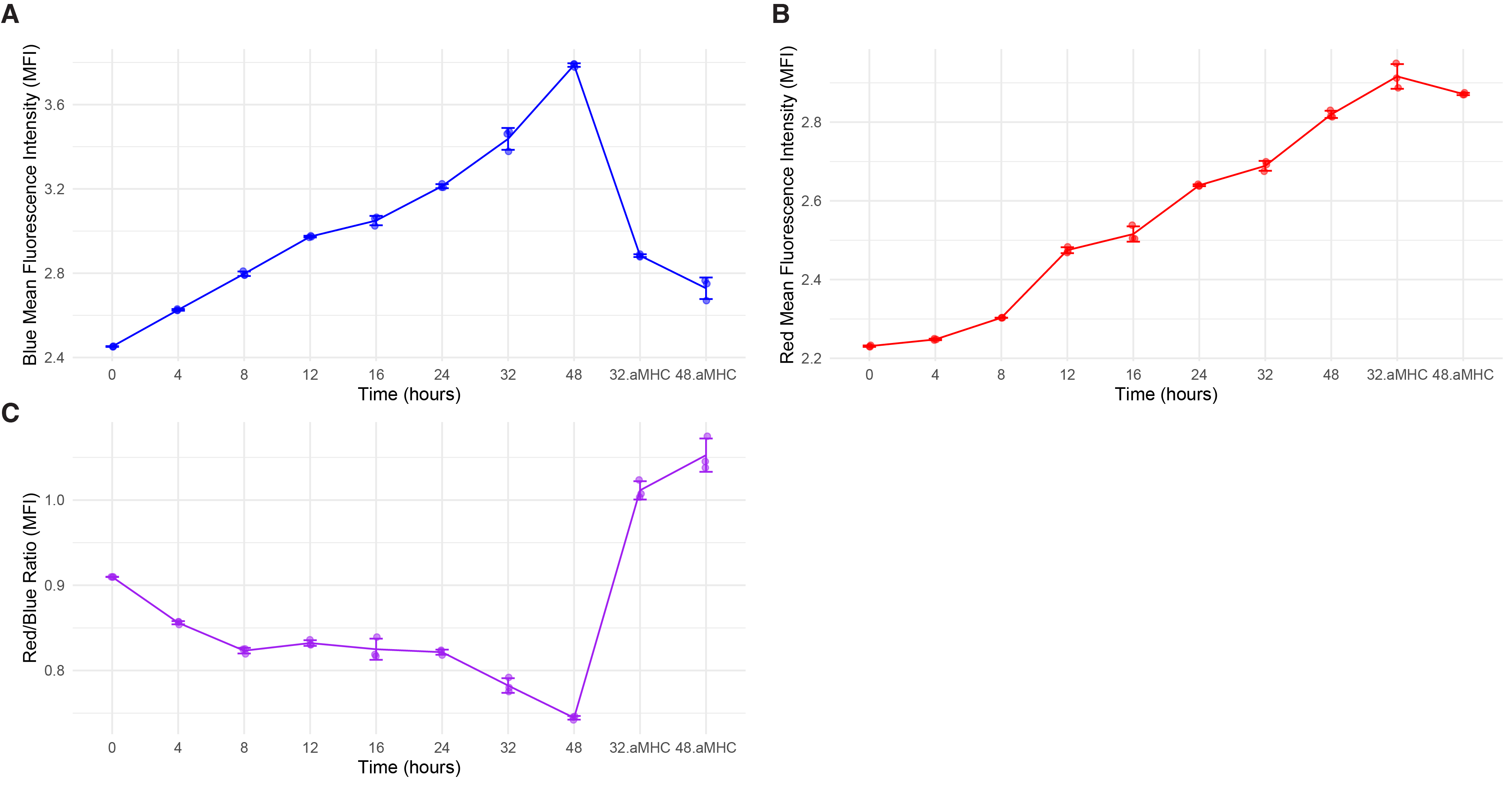}
  \caption{\textbf{Limitations of Common Approach to Fluorescent Timer Data using Mean Fluorescence Intensities (MFI).} Visualization of the Nr4a3 Tocky dataset showing (A) MFI Blue, (B) MFI Red, and (C) the ratio of Red to Blue MFI over time.}
  \label{fig:MFI}  
\end{figure}

Initially, we examined the traditional approach to Fluorescent Timer data, which relies on the mean fluorescence intensities (MFI) of Timer Blue and Red fluorescence\cite{Kisielow2019, Hernandez2013, Barry2015, Troscher2019, Peter2024, Elliot2021, Tempo2022}. This approach was applied to the Nr4a3 Tocky dataset (Figure~\ref{fig:MFI}). We observed that the MFI for both Blue and Red fluorescence increased following antigen stimulation. However, while the removal of antigen stimulation resulted in a decrease in Blue MFI, the Red MFI remained unaffected. Consequently, the ratio of Blue to Red MFI steadily declined up to 48 hours post-stimulation but surged upon cessation of stimulation.

This approach revealed significant limitations: 
\begin{enumerate}
    \item Comparing Timer Blue and Red fluorescence directly is problematic without adequate normalization and preprocessing, making it difficult to draw reliable conclusions from raw MFI data.
    \item Consequently, interpretations based on the Red-to-Blue MFI ratio are inherently limited, as they fail to account for the underlying biological variances and the effects of antigen presence or absence.
\end{enumerate}

\subsection*{Application of the Tocky Approach: Time Course Analysis of Activated T-cells Derived from Nr4a3-Tocky Mice}

\begin{figure}[H]
  \centering
  \includegraphics[width=1\textwidth]{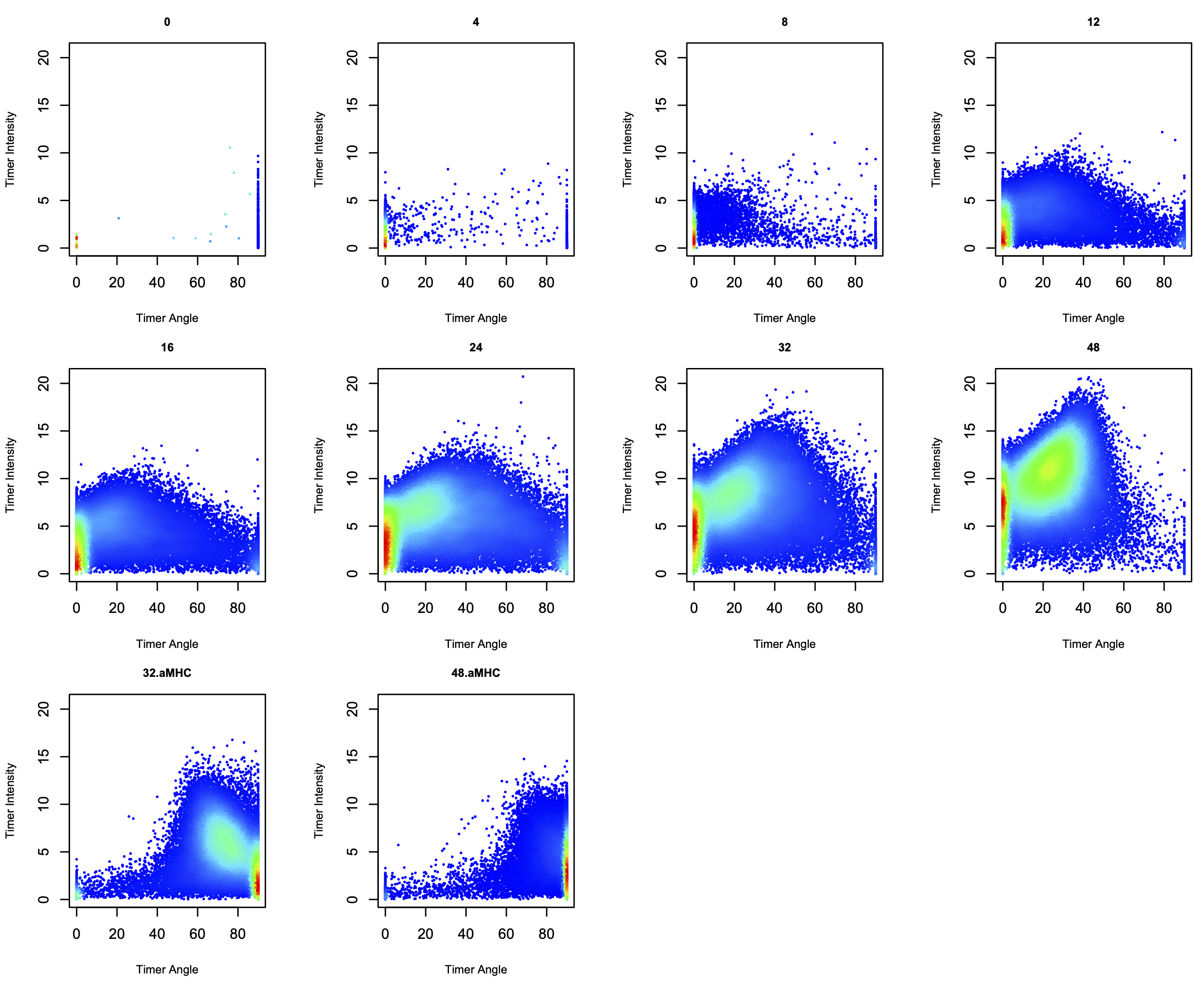}
  \caption{\textbf{Timer Angle and Intensity in Activated T-cells.} The \texttt{TockyPrep} package was utilized to normalize and transform Timer fluorescence data into Timer Angle and Timer Intensity. The function \texttt{plot\_tocky} from \texttt{TockyPrep} was employed to generate these plots. Refer to Table~\ref{tab:Table1} for sample definitions.}
  \label{fig:Plot_tocky_angle}  
\end{figure}

We then applied the Tocky approach to analyze the Nr4a3 dataset. The data preprocessing was handled by the \texttt{timer\_transform} function within the \texttt{TockyPrep} package, which includes steps such as Timer Thresholding, Timer Fluorescence Normalization, and Trigonometric Transformation. These steps effectively converted the Timer fluorescence data into quantifiable Timer Angle and Timer Intensity metrics, which are crucial for further analysis (Figure~\ref{fig:Plot_tocky_angle}).

Despite these advancements, challenges remain in how to analyze the transformed data effectively, particularly in terms of interpreting and utilizing these metrics for biological insights.

\subsection*{Strengths and Limitations of Density Plots}

\begin{figure}[H]
  \centering
  \includegraphics[width=0.75\textwidth]{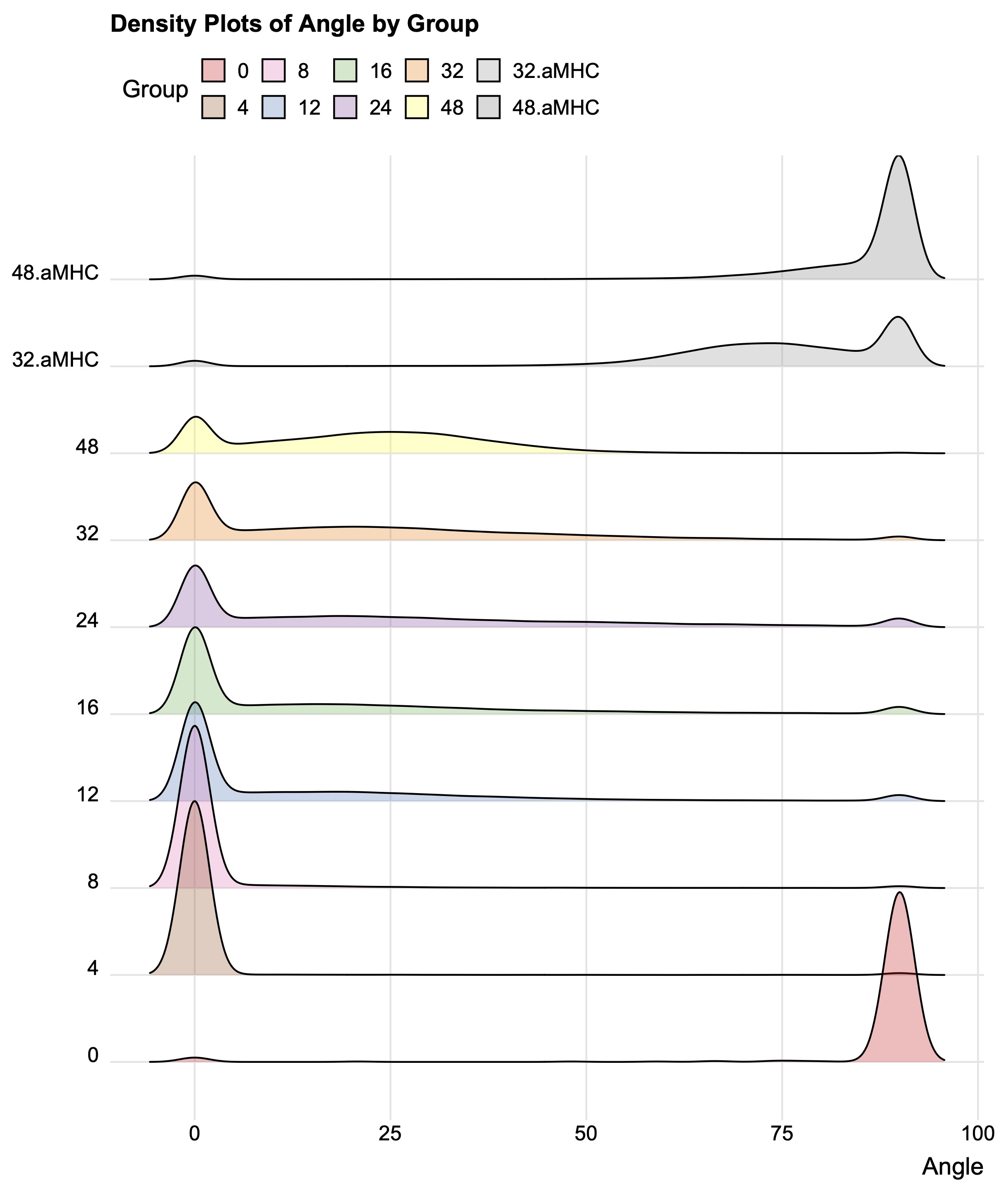}
  \caption{\textbf{Density plot of Timer Angle using the function \texttt{plotAngleDensity}.} The plot visualizes the distribution of Timer Angles within the Nr4a3 dataset, highlighting major concentrations and transitions.}
  \label{fig:plotAngleDensity}
\end{figure}

Initially, we assessed the effectiveness of density plots for visualizing and analyzing the data. While density plots provide a visual representation of the dynamics of Timer Angle progression, they are limited in their ability to offer quantitative insights. Such visualizations illustrate trends and distributions but do not replace quantitative analysis methods that can offer more definitive conclusions (Figure~\ref{fig:plotAngleDensity}).

Note that the time point 0 sample is dominated by Timer Angle \(90^\circ\) cells (Figure~\ref{fig:plotAngleDensity}). However, this is a misleading representation, as the number of these cells is small (Figures~\ref{fig:Plot_tocky_angle}. The naturally occurring Timer+ cells in Nr4a3-Tocky mice is due to the accumulation of memory-phenotype T-cells that recognised self antigens in vivo \cite{OnoSatou2024}\cite{Bending2018JCB}.

Furthermore, the inherent nature of density calculations in the plots suggests a pronounced clustering of cells at angles of 0 and 90 degrees, with apparent sparsity of transitioning cells between these values. Subsequent analyses will demonstrate that this perception is misleading, as discussed in the following sections.

\subsection*{Capturing Timer Angle Dynamics with the Tocky Locus Approach}

\begin{figure}[H]
  \centering
  \includegraphics[width=\textwidth]{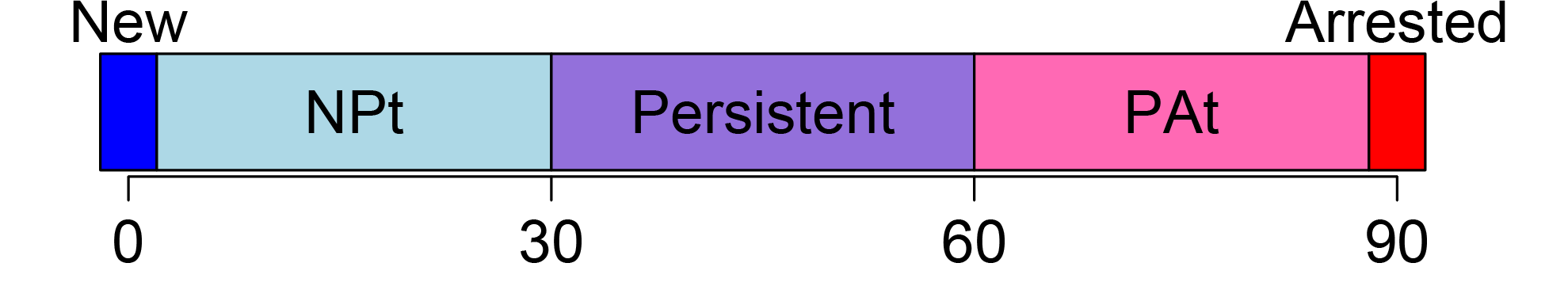}
  \caption{\textbf{The Five Tocky Loci Approach.} The \texttt{TockyLocus} package categorizes Timer fluorescence data into five distinct loci based on Timer Angles: \textbf{New} (Timer Angle = 0), \textbf{New-to-Persistent Transitioning} (NPt, \(0 < \text{Timer Angle} \leq 30\)), \textbf{Persistent} (\(30 < \text{Timer Angle} \leq 60\)), \textbf{Persistent-to-Arrested Transitioning} (PAt, \(60 < \text{Timer Angle} < 90\)), and \textbf{Arrested} (Timer Angle = 90).}
  \label{fig:PlotTockyLocusLegend}  
\end{figure}

Next, we applied the Tocky Locus approach to the Nr4a3 Tocky dataset. Timer Angle data were categorized into five Tocky loci: New, NP-t, Persistent, PA-t, and Arrested (Figure~\ref{fig:PlotTockyLocusLegend}).

\subsubsection*{Visualization of Tocky Locus in Flow Cytometric Timer Plots for Quality Control:}
\begin{figure}[H]
  \centering
  \includegraphics[width=\textwidth]{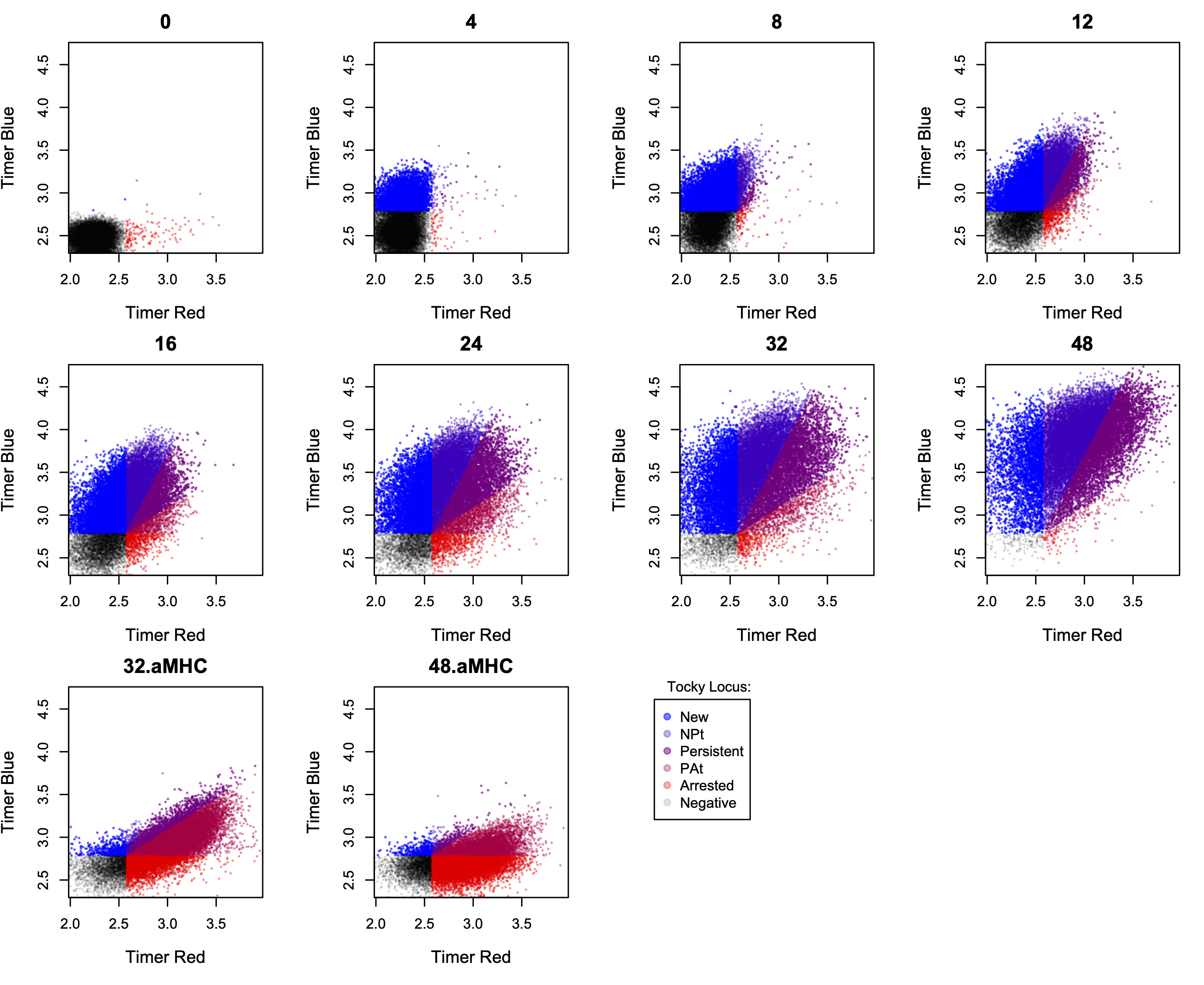}
  \caption{\textbf{Tocky Locus Plot for QC.} The Nr4a3 Tocky Dataset was analyzed using the \texttt{plot\_tocky\_locus} function from the \texttt{TockyLocus} package. Timer fluorescence data were categorized into the five Tocky loci. The percentage of cells among CD4+ T cells in each Tocky Locus is shown, demonstrating the distribution across different loci. See Table~\ref{tab:Table1} for sample definitions.}
  \label{fig:plot_tocky_locus}  
\end{figure}

For quality control purposes, the \texttt{TockyLocus} package facilitates the visualization of Tocky Locus categorization in flow cytometric plots displaying Timer blue and red fluorescence (Figure~\ref{fig:plot_tocky_locus}). These plots successfully identify Timer autofluorescence (shown in grey) and  the five Tocky loci, which are demonstrated using distinct colors in Figure~\ref{fig:plot_tocky_locus}.

\subsubsection*{Quantification of Timer Angle Dynamics by Tocky Locus Plot:}
\begin{figure}[H]
  \centering
  \includegraphics[width=\textwidth]{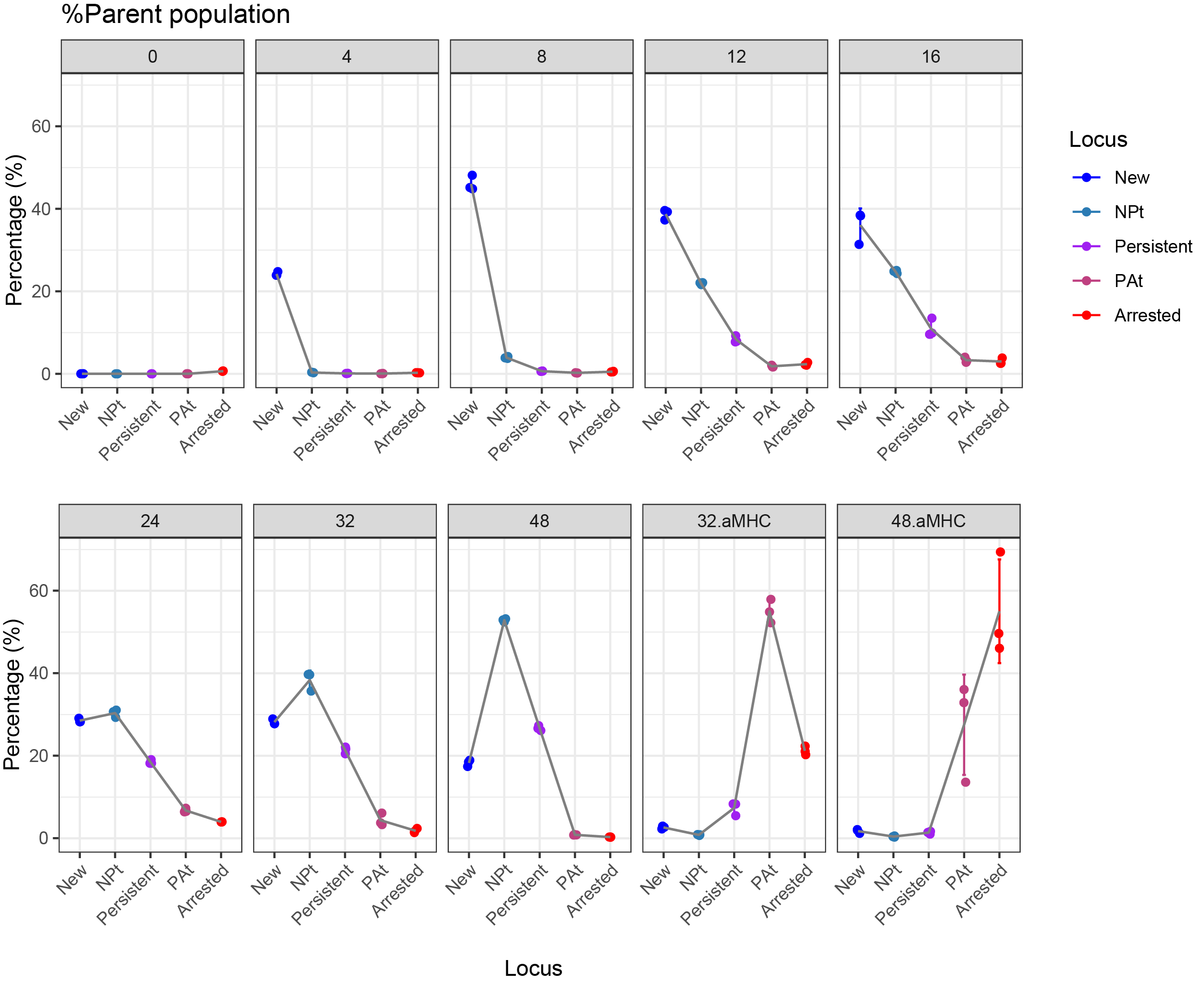}
  \caption{\textbf{Tocky Locus Plot:} This plot visualizes the dynamic progression of Tocky Loci post-antigen stimulation, illustrating how T-cells transition through various states of Timer protein expression over time. The percentage of cells among CD4+ T cells in each Tocky Locus is shown, demonstrating the distribution across different loci. See Table~\ref{tab:Table1} for sample definitions.}
  \label{fig:PlotTockyLocus}  
\end{figure}

For the quantitative analysis of Timer fluorescence dynamics, the quantification of cells within each range of Timer Angle proves to be effective. As demonstrated in Figure~\ref{fig:PlotTockyLocus}, the \texttt{PlotTockyLocus} function facilitates both the quantification and visualization of Tocky Loci in a streamlined manner. The five Tocky loci successfully capture the dynamics of Timer fluorescence, revealing that T-cells express the Timer protein as early as 4 hours post-antigen stimulation. Over time, Timer fluorescence matures and more cells are categorized into advanced Tocky Loci, including NP-t, Persistent, and PA-t. Following the termination of stimulation, there is a rapid increase in Timer Angle, transitioning the majority of cells to PA-t and subsequently to Arrested within 8 and 24 hours, respectively.

\subsection*{Optimizing the Tocky Locus Categorization}

Upon establishing the Tocky Locus approach, we proceeded to examine whether the five predefined categories adequately and effectively captured the dynamics of Timer fluorescence using the Timer Angle approach.

The number of data categories was titrated between three and seven, always including the categories 'New' and 'Arrested' as defined in Figure~\ref{fig:TockyLocus_def}. Consequently, the Timer Blue+ Red+ quadrant was subdivided into additional categories ranging from one (undivided) to five, resulting in varying degrees of data segmentation.

\begin{figure}[H]
  \centering
  \includegraphics[width=\textwidth]{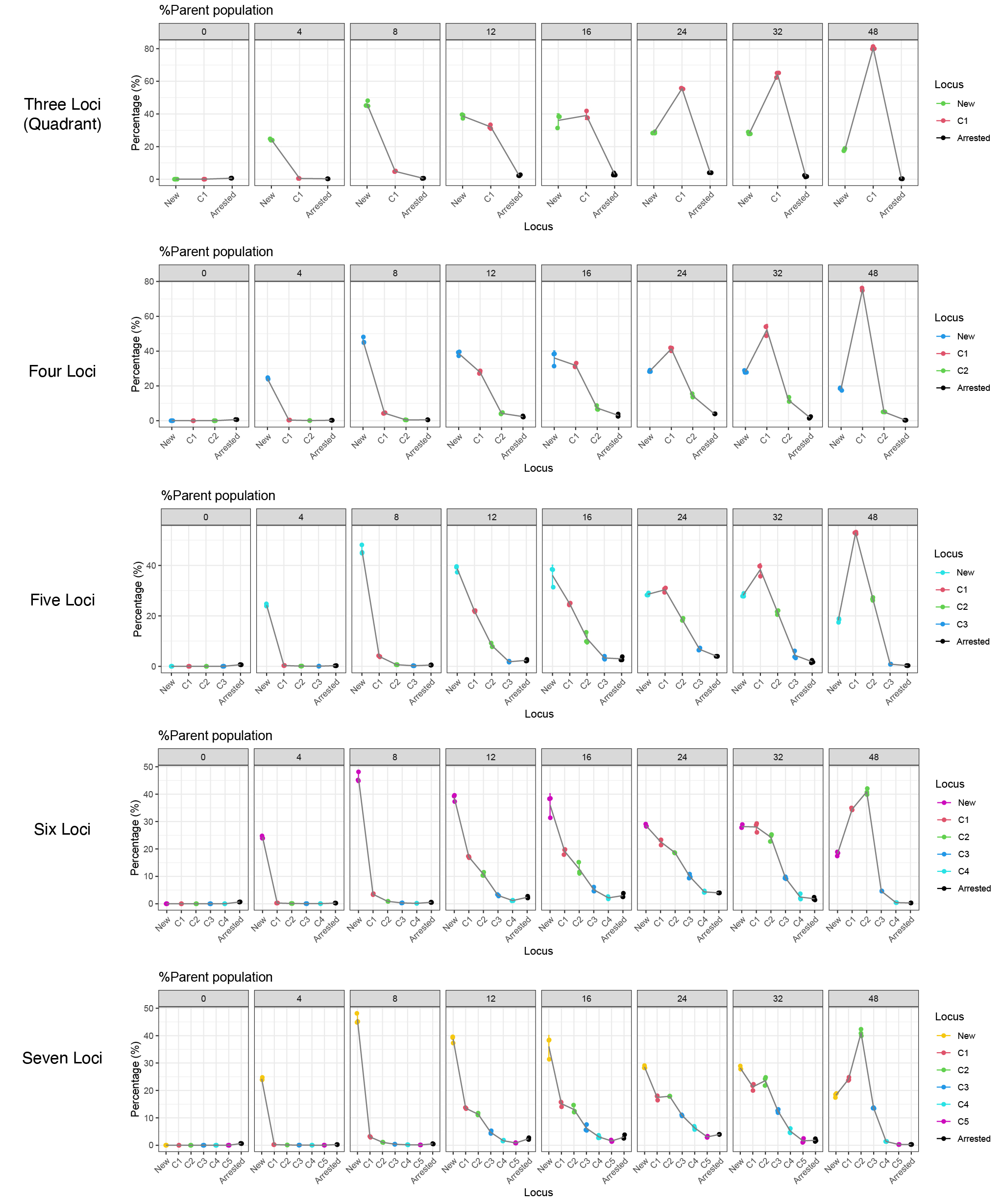}
  \caption{\textbf{Optimal Number of Tocky Loci:} This figure illustrates the impact of varying the number of Tocky Loci on the analysis of Timer Angle. Panel (A) shows Tocky Locus plots generated using different data categorization methods, highlighting how each categorization level affects the clarity and detail of the visualization.}
  \label{fig:Summary_Locus_Numbers}  
\end{figure}

Figure~\ref{fig:TockyLocus_def} shows that the three loci approach, which is equivalent to the common quadrant analysis of Timer blue and red fluorescence, is overly underpowered to capture the dynamics. In contrast, the seven-locus method may be susceptible to variations due to small numbers of cells. The locus numbers between four and six have successfully captured the nuanced dynamics of Timer Angle progression.

Given the biological significance of the Persistent locus, which is defined as the Timer Angle around \(45^\circ\), the five-locus approach was considered the most effective. This categorization not only enables the capture of cells within the Persistent locus but also maintains a relatively low number of categories, which is crucial for ensuring robustness in downstream statistical analysis.

\subsubsection*{Locus-Wise Visualization of Timer Angle Dynamics}

\begin{figure}[H]
  \centering
  \includegraphics[width=\textwidth]{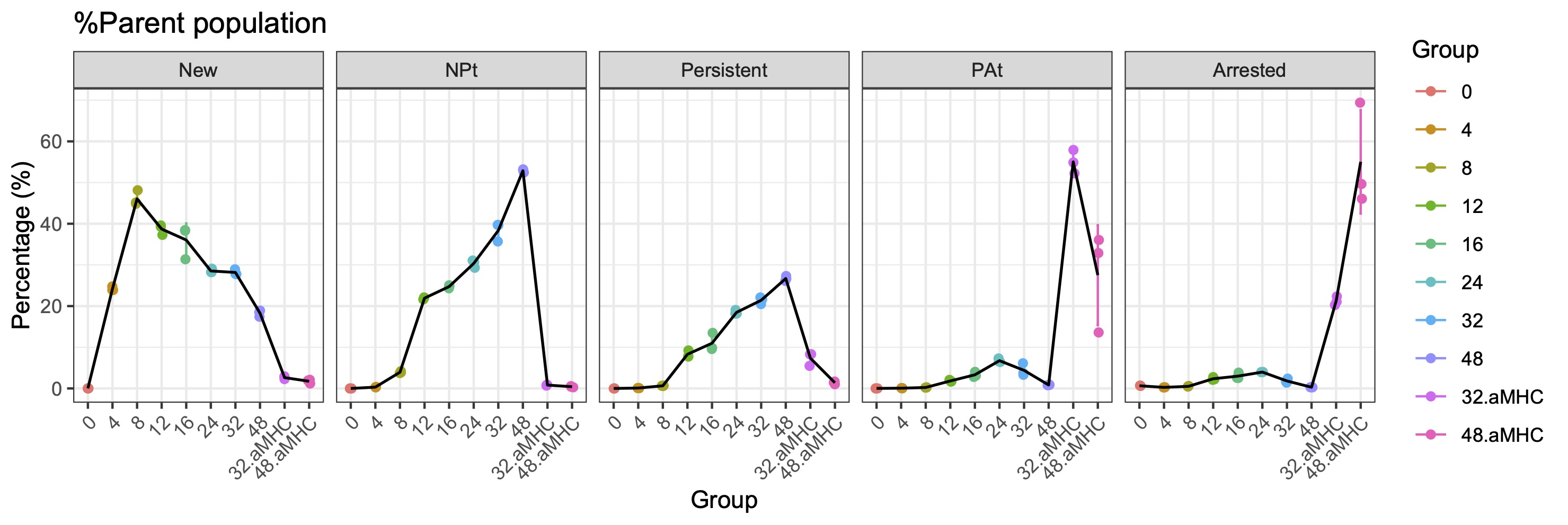}
  \caption{\textbf{Locus-Wise Visualization of Timer Angle Dynamics.} The Nr4a3 Tocky dataset was analyzed using the \texttt{PlotTockyLocus} function with the \texttt{group\_by = FALSE} option to highlight individual locus dynamics.}
  \label{fig:By_Locus}  
\end{figure}

The five-locus approach effectively facilitates the visualization of Timer Angle dynamics through locus-wise analysis (Figure~\ref{fig:By_Locus}). Given that the dataset comprises time-course data, this visualization method proves particularly useful in elucidating the cellular dynamics within each Tocky Locus.

However, the locus-wise visualization presents a drawback in that it complicates the analysis of group-specific effects (discussed below). Consequently, this approach is often considered less effective for general use, although it can be beneficial in specific scenarios where detailed, locus-specific analysis is required.

\subsection*{Effects of Percentage Metrics on Visualization}

In the previous analyses, we demonstrated the percentage of cells among CD4+ T cells within each Tocky Locus. Additionally, it is possible to calculate the percentage of cells among Timer-positive cells. This section clarifies the objectives and implications of applying these metrics.

\subsubsection*{Definitions of Percentage Metrics}
Both metrics play crucial roles in various aspects of the analysis and are defined as follows:

\begin{description}
    \item[{\%Parent}:] Denotes the proportion of cells within each locus compared to the entire parent population (e.g., all CD4+ T cells). This metric is typically preferred for most analyses.
    \begin{itemize}
        \item \textbf{Aim:} To analyze the dynamics of Timer-expressing cells within the relevant cell population.
        \item \textbf{Note:} This metric unravels the prevalence of Timer-expressing cells in each locus within the total cell population, elucidating the dynamics of Timer expression throughout the parent population. 
    \end{itemize}

    \item[{\%Timer}:] Represents the proportion of cells within each specific locus relative to the total number of Timer-positive cells. 
    \begin{itemize}
        \item \textbf{Aim:} To analyze the proportion of Timer-expressing cells in each Tocky Locus in the context of all Timer-positive cells.
        \item \textbf{Note:} This metric is particularly useful for highlighting the most significant loci within the Timer-positive cells, facilitating a focused analysis on the most active segments within Timer-expressing cells.
    \end{itemize}
\end{description}

\subsubsection*{Comparison between Percentage Metrics}

\begin{figure}[H]
  \centering
  \includegraphics[width=\textwidth]{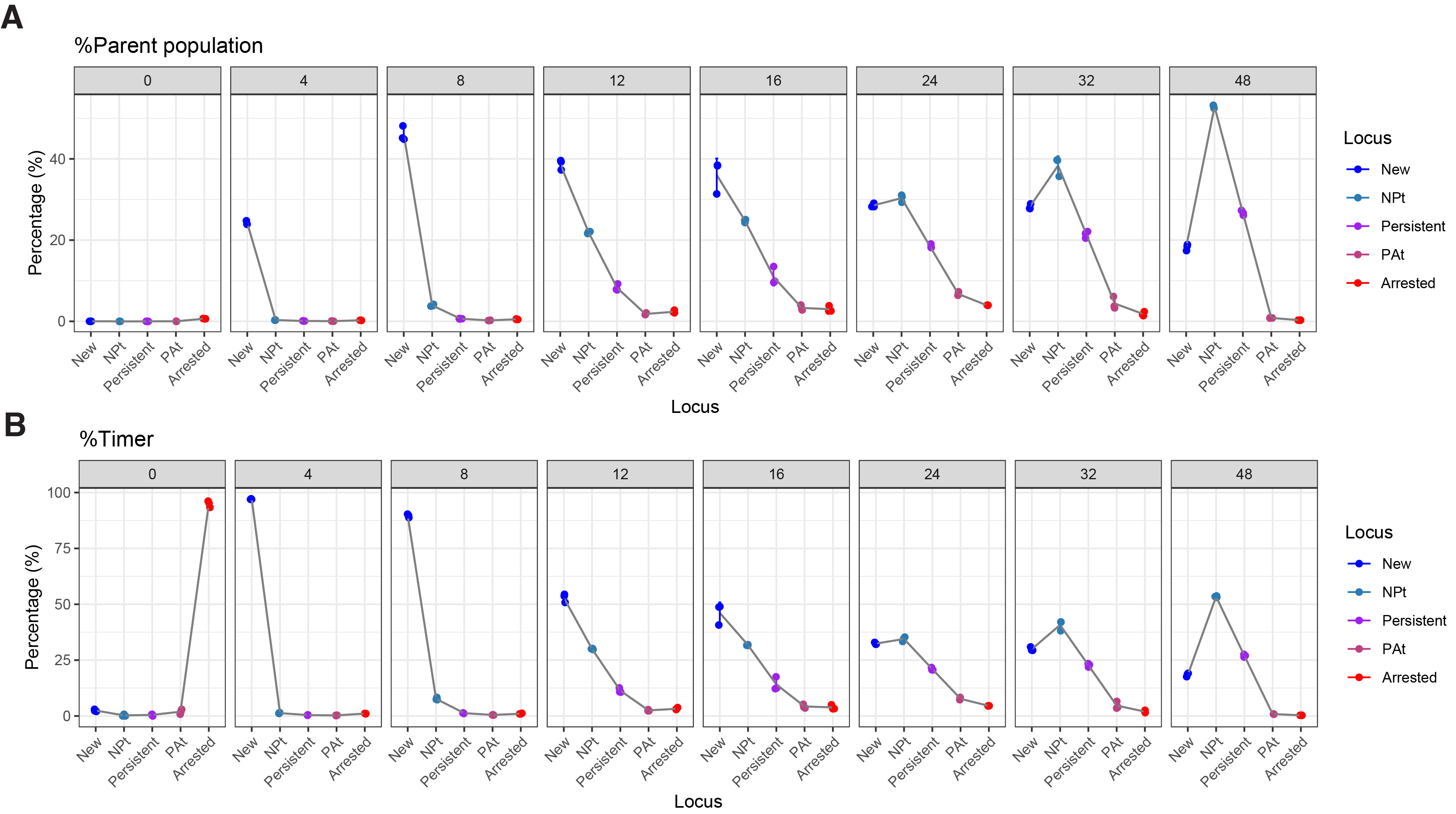}
  \caption{\textbf{Comparison of Visualization Metrics.} (A) Visualization using the Percentage-Parent metric, illustrating the proportion of cells within each locus relative to the entire parent population. (B) Visualization using the Percentage-Timer metric, highlighting the proportion of cells within each locus relative to all Timer-positive cells.}
  \label{fig:PercentParent_Timer}  
\end{figure}

Figure~\ref{fig:PercentParent_Timer} compares the visualization outcomes for the two metrics. Panel (A) displays the visualization based on the Percentage-Parent metric, while Panel (B) is produced using the Percentage-Timer metric.

Notably, the Percentage-Timer metric tends to overemphasize the accumulation of small numbers of Timer-positive cells in the Arrested locus (see above). This observation is not only from this specific analysis but from our extensive investigations, which generally affirm the importance of the Percentage-Parent metric in many scenarios.

\paragraph{Biological Conclusion:} The analysis results from the Tocky Locus approach corroborate that Fluorescent Timer dynamics within T-cells from Nr4a3 Tocky mice accurately reflect the temporal dynamics of TCR signaling. Additional details and supporting experimental results can be found in \cite{Bending2018JCB}.

\subsection*{Application of Tocky Locus to Two-Group Analysis: The OX40 Dataset from Foxp3-Tocky Mice}

Lastly, we apply the Tocky Locus approach to the OX40 dataset, which analyzed two groups. Briefly, Foxp3 Tocky mice were sensitized on their abdominal skin by a skin allergen (precisely, a hapten called Oxazolone). Three mice were treated by anti-OX40 antibody, which enhances OX40 signaling, while four mice were treated by isotype antibody as control. Five days after the sensitization, the mice were challenged on their ear skin with a small amount of Oxazaolone. The mice were analyzed at 96 hours after the challenge, and skin-infiltrating T-cells were analyzed by flow cytometry for Fluorescent Timer profiles\cite{Bending2018EMBO}.

\subsubsection*{Strengths and Limitations of Density Plots Using the OX40 Dataset}

\begin{figure}[H]
  \centering
  \includegraphics[width=0.7\textwidth]{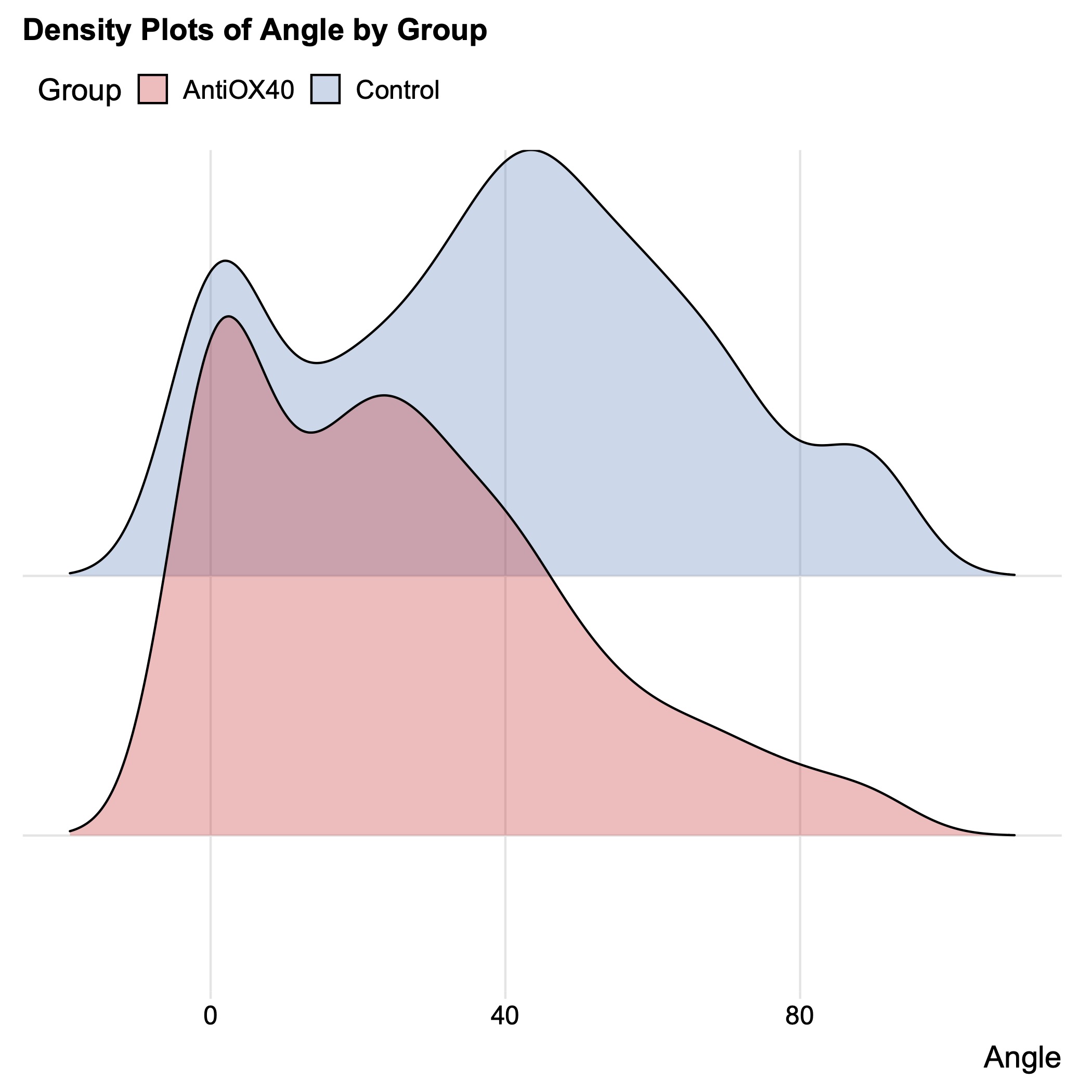}
  \caption{\textbf{Density Plot Visualization.} This figure demonstrates the use of density plots in the context of the \texttt{percentTimer} option, highlighting both their strengths and limitations in representing the data.}
  \label{fig:DensityPlotOX40}  
\end{figure}

Figure~\ref{fig:DensityPlotOX40} illustrates the strengths of density plots in visualizing the overall tendency of data, showing the potential decrease of Timer Angle in the Anti-OX40 group. However, this approach clearly has limitations due to the lack of quantitative measures.

\subsubsection*{Tocky Locus Approach to Therapeutic Effects of Anti-OX40 Antibody}

Using the OX40 dataset, we explored whether and how the Tocky Locus approach facilitates quantitative analysis of Timer profiles. To this end, we varied the number of data categories, or Tocky loci, and analyzed cellular dynamics using the \texttt{PlotTockyLocus} function of the \texttt{TockyLocus} package.

\begin{figure}[H]
  \centering
  \includegraphics[width=\textwidth]{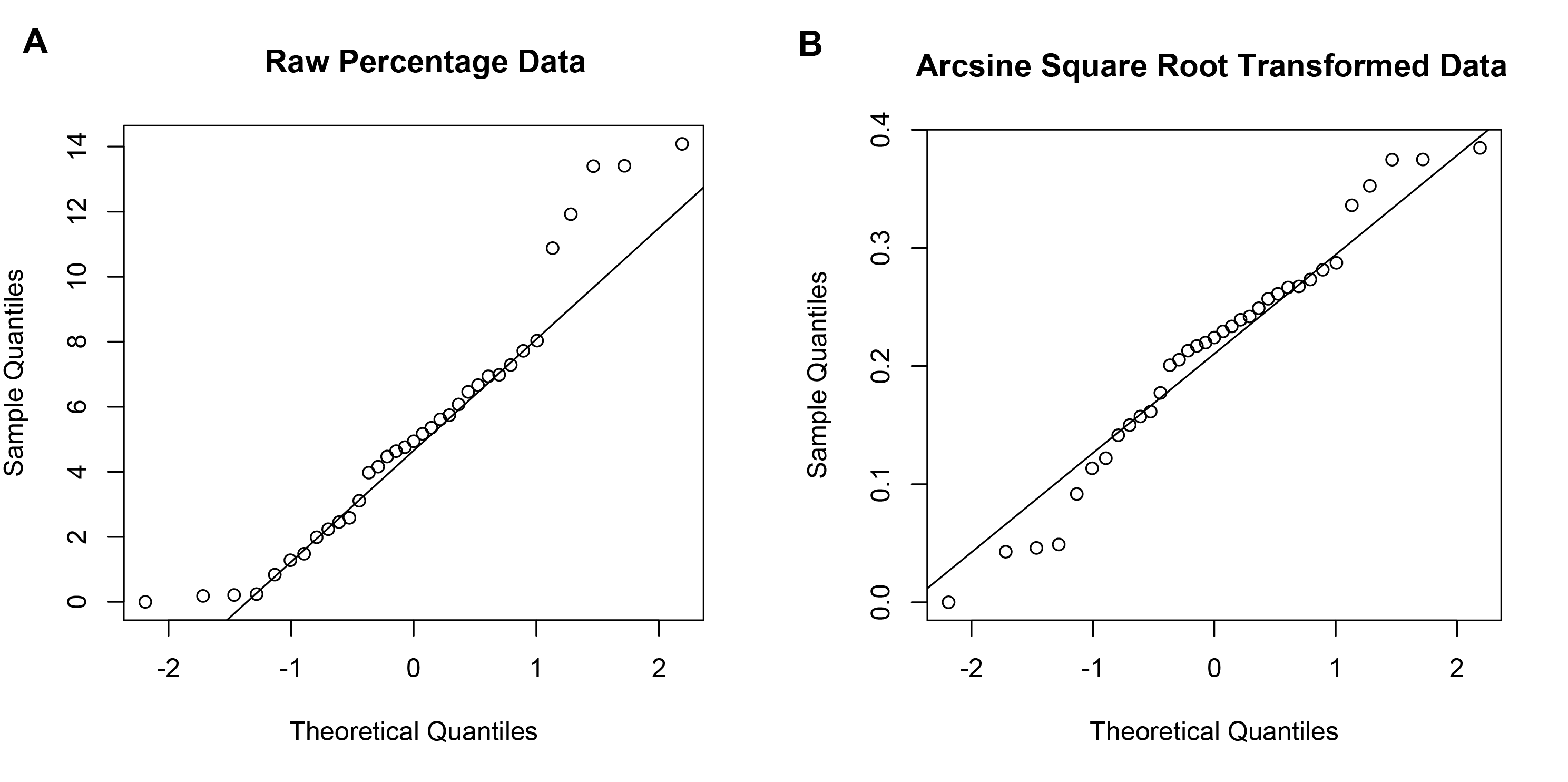}
  \caption{\textbf{Effect of Data Transformation.} This figure shows the QQ plot for normality assessment after applying the arcsine square root transformation to the data.}
  \label{fig:qqnorm}  
\end{figure}

Statistical analyses were conducted using data transformed by the Arcsine Square Root method. The transformed data were evaluated for normality using the Shapiro-Wilk test, which confirmed their normal distribution (Figure~\ref{fig:qqnorm}). This was followed by a t-test with p-value adjustments using the False Discovery Rate (FDR) to address multiple testing issues across the five loci.

\begin{figure}[H]
  \centering
  \includegraphics[width=\textwidth]{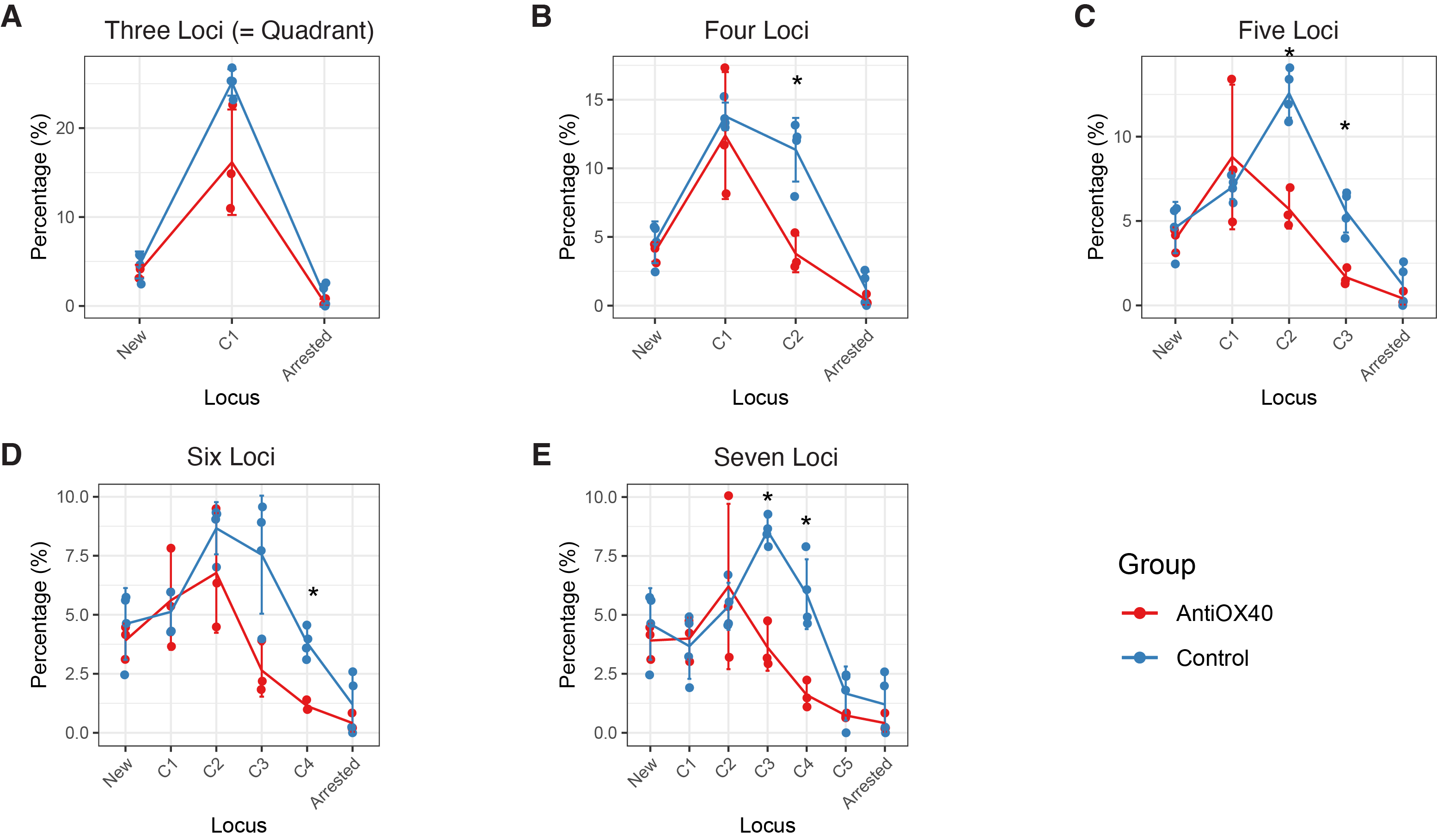}
  \caption{\textbf{Effect of Varying Tocky Loci Numbers.} This figure displays the impact of using different numbers of loci on the analysis. Panels A, B, C, D, and E show results from analyses using three, four, five, six, and seven loci, respectively. Asterisks indicate significant results with an adjusted p-value $< 0.05$, using arcsine square root data transformation, normality tests, and t-statistics.}
  \label{fig:OX40_locus_num}  
\end{figure}

Figure~\ref{fig:OX40_locus_num} presents the analysis with varying numbers of Tocky loci. The five-loci approach effectively captured the dynamics of Timer Angle, revealing significant group-specific effects in C2 (Persistent) and C3 (PA-t) categories. Conversely, using three and six loci proved less effective in identifying these effects, highlighting the optimal utility of the five-loci configuration for statistical robustness.

\paragraph{Biological Conclusion:} The analysis results from the five-loci approach indicate that the anti-OX40 antibody significantly reduces the population of CD4+ T-cells that persistently transcribe the Foxp3 gene. Our previous work has detailed the characteristics of these CD4+ T-cells, identifying them as activated regulatory T-cells that express high levels of OX40 as well as inhibitory co-receptors, including CTLA-4 and ICOS \cite{Bending2018EMBO}.

\section*{Discussion}

The major challenge in the use of Fluorescent TImer data is in the data analysis. Currently, it is prevalent to use MFI methods for analyzing Fluorescent Timer data  \cite{Kisielow2019, Hernandez2013, Barry2015, Troscher2019, Peter2024, Elliot2021, Tempo2022}. The alternative traditional approach is a heavy reliance on arbitrary gating or the quadrant approach. The latter categorizes fluorescence into groups such as Blue+Red-, Blue+Red+, Blue-Red+, and Timer- cells \cite{Bortoluzzi2021, Elliot2021, Tempo2022}. However, our investigations have demonstrated that these approaches are underpowered for both visualization and statistical analysis.

In contrast, the Tocky Locus approach, employing a five-locus system, offers a more nuanced analysis by accurately capturing biologically important cells with specific Timer Angles, such as those at approximately \(45^\circ\) in the Persistent locus, indicative of sustained transcriptional activities. However, categorizing into too many loci can lead to issues with small sample sizes in each category. The challenge has been to balance granularity with practical analysis constraints, which has led to the adoption of the five-locus system as a standard \cite{Bending2018JCB, Bending2018EMBO, Hassan2022} (Figure~\ref{fig:PlotTockyLocusLegend}).

Meanwhile, to date, our group has conducted several studies that successfully utilized the Tocky Locus approach with five loci, elucidating the dynamics of reactive T-cells during development, homeostasis \cite{Bending2018JCB}, inflammation \cite{Bending2018EMBO}, and cancer \cite{Hassan2022}. However, the Tocky Locus approach had not been formally examined before the current study, which is the first to thoroughly elaborate on the Tocky Locus approach and provides robust evidence supporting this data categorization method for understanding flow cytometric Timer fluorescence data.

The current implementation incorporates statistical methods specifically tailored for analyzing percentage data within each Tocky Locus. Utilizing the five-locus approach, it is advisable to apply a p-value adjustment method to address multiple testing issues effectively. However, analyzing percentage data, particularly with small sample sizes, poses significant challenges. In response to these challenges, we have expanded our statistical toolkit to include not only the non-parametric Mann-Whitney test followed by p-value adjustment but also data transformation techniques such as the Arcsine Square Root and Logit transformations. These transformations are prerequisites for applying the parametric Student's t-test. The selection of a data transformation method depends on the specific requirements of the case and the data structure. These methods are designed to enhance the effectiveness and precision of the Tocky Locus approach.

\section*{Conclusion}

In this study, we have developed a data categorization method using five Tocky Loci, based on biological significance observed in experimental data and theoretical investigations. The \texttt{TockyLocus} package, equipped with visualization and statistical tools, including data transformation methods, is designed to support research using flow cytometry and Fluorescent Timer reporter systems. These methods are expected to improve the quantitative analysis of Timer fluorescence data, providing researchers with more accurate and detailed analytical capabilities. We anticipate that this toolkit will advance understanding of dynamic cellular processes across various fields of molecular and cell biology, including immunology.

\section*{Acknowledgements}

MO was supported by a CRUK Programme Foundation Award (DCRPGF-100007), an MRC grant (MR/S000208/1), and a BBSRC David Phillips Fellowship (BB/J013951/2).

\section*{Code Availability}
The \texttt{TockyLocus} package is freely available at our GitHub site: \url{https://github.com/MonoTockyLab/TockyLocus}. Documentations can be accessed at \url{https://MonoTockyLab.github.io/TockyLocus}.

\bibliographystyle{unsrt}
\bibliography{TockyLocus}

\end{document}